\documentclass[journal,10pt]{IEEEtran} 
\usepackage{amssymb}
\usepackage{amsmath}
\usepackage{bm}
\usepackage{amsfonts}
\usepackage{graphicx}
\usepackage{subfigure}
\usepackage{cite}
\usepackage{enumerate}
\usepackage{url}
\usepackage{setspace}
\usepackage[ruled,linesnumbered]{algorithm2e}
\usepackage{makecell}
\usepackage{mathtools}
\usepackage{epstopdf}
\usepackage{dsfont}
\usepackage[usenames,dvipsnames]{color}
\usepackage{colortbl}
\definecolor{tianlan}{rgb}{.94, 1, 1}
\definecolor{qianhuilan}{rgb}{.69,.88,.90}
\definecolor{xiangyabai}{rgb}{.98, 1,.94}
\definecolor{baixinren}{rgb}{1, .92, .80}
\definecolor{danke}{rgb}{.99, .90, .79}
\definecolor{milucheng}{rgb}{.94, 1, .94}
\definecolor{huang1}{rgb}{1, .98, .84}
\definecolor{lan1}{rgb}{.88, .92, .98}
\usepackage{makecell}
\usepackage{multirow}
\usepackage[table,xcdraw]{xcolor}
\usepackage{makecell}
\allowdisplaybreaks[4]
\begin{document}

\newtheorem{definition}{\bf ~~Definition}
\newtheorem{observation}{\bf ~~Observation}
\newtheorem{theorem}{\bf ~~Theorem}
\newtheorem{proposition}{\bf ~~Proposition}
\newtheorem{remark}{\bf ~~Remark}

\title{Reconfigurable Holographic Surfaces for Future Wireless Communications}
\author{
\IEEEauthorblockN{
Ruoqi Deng,~\IEEEmembership{Graduate Student Member,~IEEE},
Boya Di,~\IEEEmembership{Member,~IEEE},
Hongliang Zhang,~\IEEEmembership{Member,~IEEE},\\
Dusit Niyato,~\IEEEmembership{Fellow,~IEEE},
Zhu Han,~\IEEEmembership{Fellow,~IEEE},
H. Vincent Poor,~\IEEEmembership{Life Fellow,~IEEE},\\
and Lingyang Song,~\IEEEmembership{Fellow,~IEEE}}

\thanks{This work has been submitted to the IEEE for possible publication. Copyright may be transferred without notice, after which this version may no longer be accessible.}
\thanks{Ruoqi Deng, Boya Di, and Lingyang Song are with Department of Electronics, Peking University, Beijing, China (email: ruoqi.deng@pku.edu.cn; diboya92@gmail.com; lingyang.song@pku.edu.cn).}
\thanks{Hongliang Zhang and H. Vincent Poor are with Department of Electrical Engineering, Princeton University, NJ, USA (email: hongliang.zhang92@gmail.com).}
\thanks{Dusit Niyato is with the School of Computer Science and Engineering, NTU, Singapore 639798 (Email: dniyato@ntu.edu.sg).}
\thanks{Zhu Han is with Electrical and Computer Engineering Department, University of Houston, Houston, TX, USA, and also with the Department of
Computer Science and Engineering, Kyung Hee University, Seoul, South Korea (Email: zhuhan22@gmail.com).}
\vspace{-0.6cm}
}
\maketitle

\begin{abstract}
Future wireless communications look forward to constructing a ubiquitous intelligent information network with high data rates through cost-efficient devices. Benefiting from the tunability and programmability of metamaterials, the reconfigurable holographic surface (RHS) composed of numerous metamaterial radiation elements is developed as a promising solution to fulfill such challenging visions. The RHS is more likely to serve as an ultra-thin and lightweight surface antenna integrated with the transceiver to generate beams with desirable directions by leveraging the holographic principle. This is different from reconfigurable intelligent surfaces (RISs) widely used as passive relays due to the reflection characteristic. In this article, we investigate RHS-aided wireless communications. Starting with a basic introduction of the RHS including its hardware structure, holographic principle, and fabrication methodologies, we propose a hybrid beamforming scheme for RHS-aided multi-user communication systems. A joint sum-rate maximization algorithm is then developed where the digital beamforming performed at the base station and the holographic beamforming performed at the RHS are optimized iteratively. Furthermore, key challenges in RHS-aided wireless communications are also discussed.

\end{abstract}




\section{Introduction}
To enable a ubiquitous intelligent information network, the forthcoming sixth generation (6G) wireless communications put stringent requirements on antenna technologies such as accurate beam steering and capacity enhancement  \cite{C-2020}. Though massive multiple-input multiple-output (MIMO) technology relying on large-scale phased arrays is perceived to achieve these goals, inherent limitations of phased arrays severely hinder their future development. Specifically, phased arrays highly rely on power amplifiers consuming high power and numerous phase shifters, which are costly especially in high-frequency bands. Therefore, more cost-efficient antenna technologies are required to meet the exponentially increasing data demand in future wireless communications\cite{BHL-2020}.

\begin{figure}[t]
\centering
\includegraphics[width=3.5in]{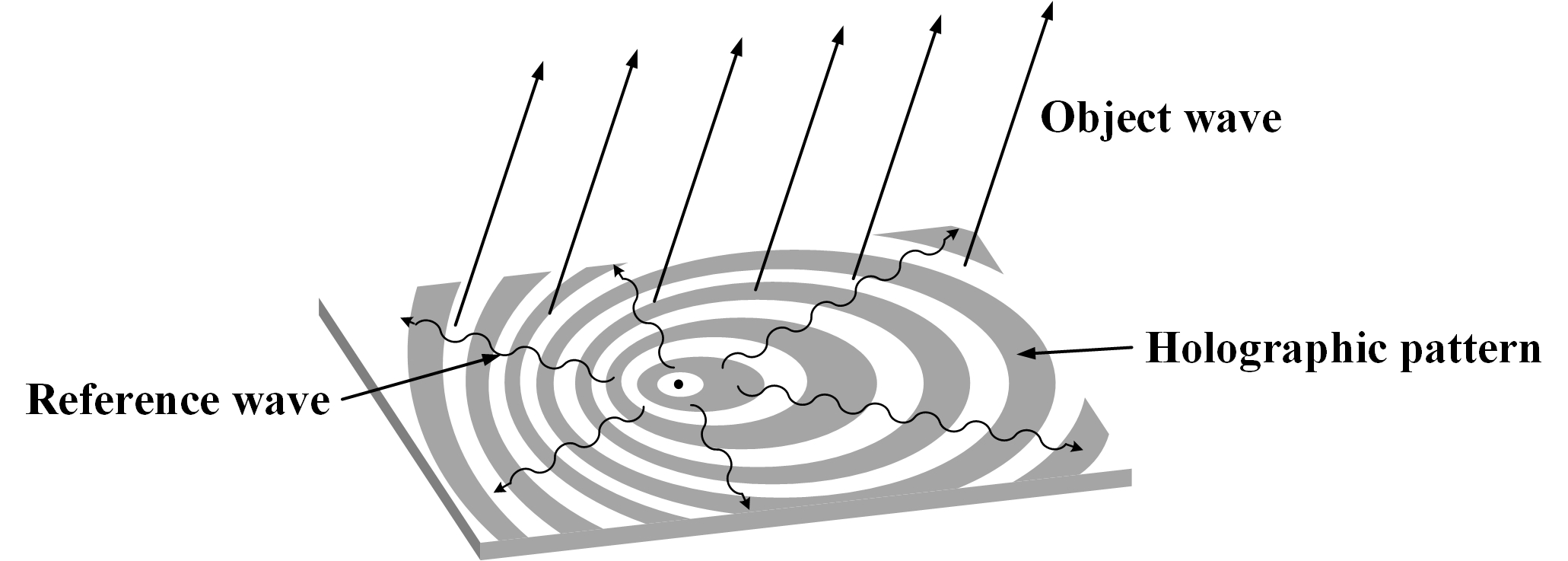}
\caption{Illustration of the holographic antenna \cite{BJJ-2010}.}
\label{URHS}
\end{figure}

To overcome the deficiency of existing antenna technologies, holographic antennas, as small-size and low-cost planar antennas, have attracted increasing attention due to their capability of multi-beam steering with low power consumption. As shown in Fig. \ref{URHS}, the holographic antenna utilizes meta patches to construct a \emph{holographic pattern} on the surface, which records the interference between the incident electromagnetic wave generated by the holographic antenna (which is also called reference wave) and the desired object wave based on the holographic interference principle \cite{BJJ-2010}. When the reference wave propagates on the antenna surface, its radiation characteristics can be changed by the holographic pattern to generate the desired radiation pattern. However, the conventional holographic antenna is not applicable to complex and unstable wireless environments since its radiation pattern is fixed once the holographic pattern is constructed.

Fortunately, benefiting from the tunability and programmability of the metamaterial, the emerging reconfigurable holographic surface (RHS) shows great potential to break through the bottleneck of the conventional holographic antenna \cite{OD-2017}. An RHS is an ultra-thin and lightweight surface antenna inlaid with numerous metamaterial radiation elements based on the printed-circuit-board (PCB) technology. By controlling the electromagnetic response of the metamaterial, the holographic pattern and the corresponding desired beam directions can be easily reconfigured. Specifically, each element can control the \emph{radiation amplitude} of the reference wave electrically to generate the desired beams according to the holographic pattern. Therefore, without heavy mechanics and complex phase-shifting circuits, the RHS can achieve dynamic beamforming, and such a beamforming technique is also known as \emph{holographic beamforming}.

Apart from RHSs, the recent development of meta-surfaces also introduces another hardware  technology for wireless communication enhancement called reconfigurable intelligent surfaces (RISs) \cite{M-2019}. Specifically, an RIS is also an ultra-thin surface containing multiple metamaterial elements with controllable electromagnetic properties. It can reflect the incident signals and generate directional beams towards receivers directly. RISs can create an emerging paradigm of smart radio environments \cite{M-2019}, and the hybrid beamforming scheme for RIS-based multi-user communications has also been developed \cite{BHL-2020}. Though RHSs and RISs are both metasurfaces capable of beamforming, they are different in several aspects as listed below.

\begin{itemize}
\item \emph{Physical Structure}: The RHS's RF front end is integrated into a PCB, enabling a convenient implementation at the transceiver. Therefore, no extra control link is required for the RHS to construct the holographic pattern. In contrast, the RF front end of the RIS is on the outside of the meta-surface due to its reflection characteristic. An extra control link is required between the RIS and the transmitter to reconstruct the phase shifts \cite{BHL-2020}.
\item \emph{Operating Mechanism}: The RHS is a leaky-wave antenna. It utilizes the method of series feeding where radiation elements are located progressively farther and farther away from the feed point. The reference wave generated by the feed propagates on the RHS and excites the radiation element one by one. In contrast, the RIS is a reflection antenna. It utilizes the method of parallel feeding where all radiation element is excited by the incident signals at the same time.
\item \emph{Typical Applications}: The RHS is more likely to serve as transmit and receive antennas, which can be easily mounted on mobile platforms to provide high-throughput connectivity due to its high integration and ultra-thin structure. It can also be integrated with the transceiver of a radar for localization or imaging. The RIS is widely used as a passive relay \cite{CBH-2021}. One typical application scenario of RISs is to be deployed in the cell edge for the cell coverage extension and the cell-edge users' performance improvement.
\end{itemize}

In the literature, research on the RHS from both academia and industry focuses on the hardware component design \cite{OD-2017} and radiation pattern control \cite{MSJ-2014}. In \cite{OD-2017}, an RHS capable of generating multi-beam radiation patterns has been presented. In \cite{MSJ-2014}, a radiation pattern control algorithm for RHSs has been developed for sidelobe suppression. Two commercial prototypes of the RHS have been developed by Pivotal Commwave \cite{Pi} and Kymeta \cite{Ky}. Specifically, Pivotal Commwave can provide customized RHS systems between 1 GHz and 70 GHz. Kymeta has introduced the world's first commercial RHS for electronic scanning in satellite communications.

However, most existing works only demonstrate the viability of the RHS to achieve dynamic multi-beam steering. None of them has studied holographic beamforming for system performance improvement in wireless communications. New challenges have arisen in holographic beamforming for sum rate maximization. On the one hand, since the traditional phase-controlled analog beamformer is a complex-valued matrix with complex-domain phase constraints, existing algorithms do not work well for amplitude-controlled holographic beamforming. A new beamforming scheme needs to be developed to handle the complex-domain optimization problem subject to unconventional real-domain amplitude constraints, coupled with the superposition of the radiation waves from different radiation elements. On the other hand, the coupling between all radiation elements simultaneously with the propagating electromagnetic wave complicates the holographic beamforming design.

In this article, we introduce the hardware structure and holographic principle of the RHS. Following that, we investigate the feasibility of applying the RHS to wireless communications from two following aspects:
\begin{itemize}
\item \emph{Fabrication Methodology and Full-Wave Analysis:} We present the fabrication methodology of the RHS and conduct a full-wave analysis of the RHS. Simulation results at system level and component level are also presented to verify that the RHS can change the beam directions by controlling the radiation amplitude of each radiation element.
\item \emph{Hybrid Beamforming Scheme:} To maximize the sum rate in the RHS-aided multi-user communication system, we propose a hybrid beamforming scheme where the digital beamforming performed at the base station (BS) and the holographic beamforming performed at the RHS are jointly optimized. The key challenges in RHS-aided wireless communications are also discussed.
\end{itemize}

The rest of this article is organized as follows. In Section \uppercase\expandafter{\romannumeral2}, we introduce the hardware structure and the holographic principle of the RHS. In Section \uppercase\expandafter{\romannumeral3}, we present the fabrication methodology and a full-wave analysis of the RHS. In Section \uppercase\expandafter{\romannumeral4}, we propose a hybrid beamforming scheme for RHS-aided wireless communications, based on which a joint sum rate maximization algorithm is developed. Key challenges in RHS-aided wireless communications are elaborated in Section \uppercase\expandafter{\romannumeral5}. Finally, we draw the conclusions in Section \uppercase\expandafter{\romannumeral6}.

\section{Hardware and Principle of Reconfigurable Holographic Surfaces}

\subsection{Hardware Structure}\label{IIA}
The RHS is a special leaky-wave antenna. The incident electromagnetic wave propagates along the guiding structure of the RHS. Modulated by the RHS, the propagating wave can be transformed into a leaky wave through discontinuities on the surface and leaks out its energy into free space for radiation \cite{OD-2017}. The superposition of the radiation waves from different discontinuities on the surface then generates the desired directional beams.

\begin{figure}[t]
\centering
\includegraphics[width=3.4in]{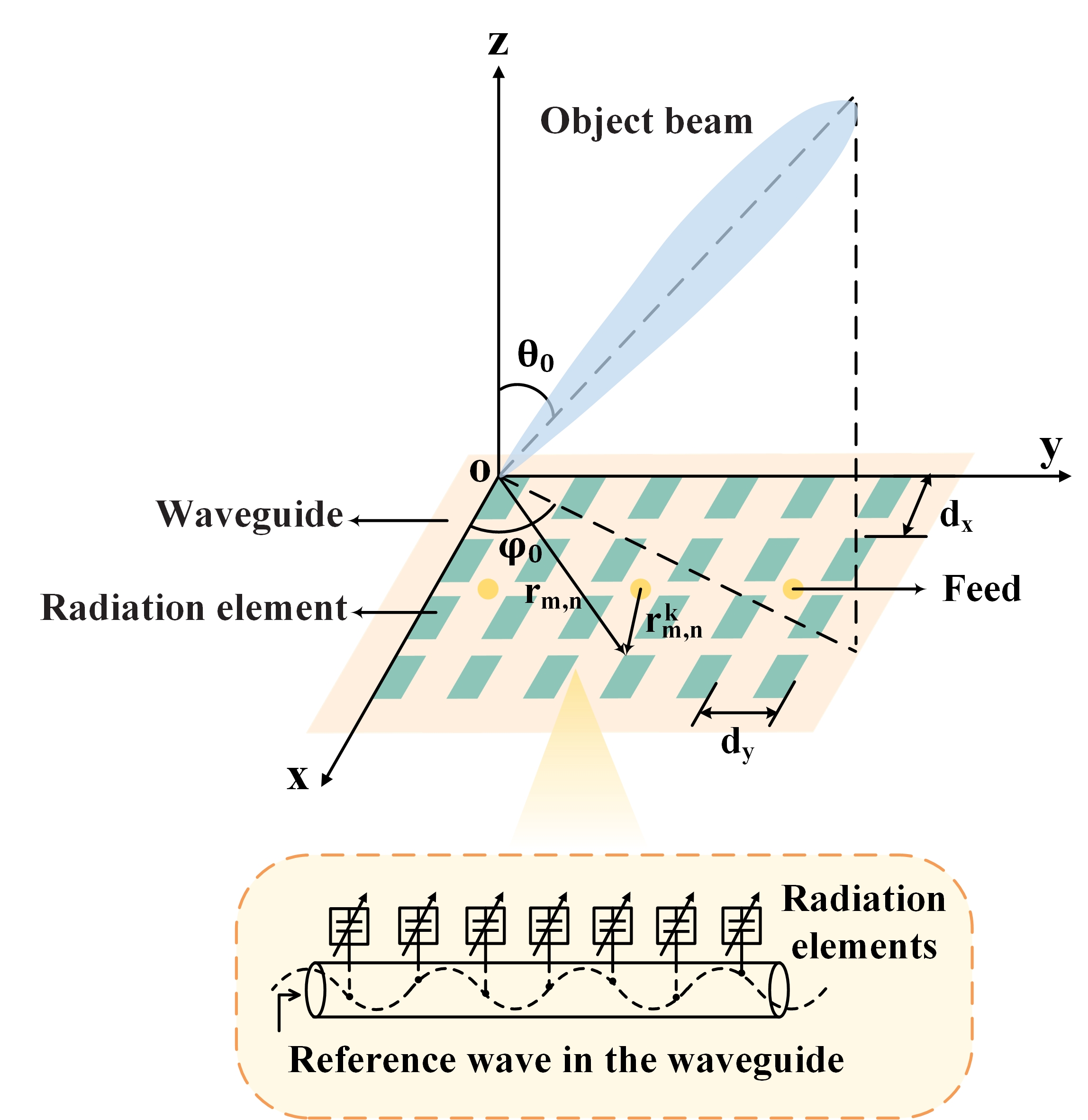}
\caption{Illustration of the RHS.}
\label{fig1}
\end{figure}

As shown in Fig. \ref{fig1}, the RHS mainly consists of three parts. Their specific functions are elaborated below.
\begin{itemize}
\item \emph{Feed}: The feeds are embedded in the bottom layer of the RHS to generate the incident electromagnetic waves, which are also called reference waves, propagating along the RHS and exciting the electromagnetic field of the RHS. Such a structure enables an ultra-thin RHS compared to traditional antenna arrays where the feeds are usually bulky and outside the antenna surface.

\item \emph{Waveguide}: The waveguide is the propagation medium of the reference wave. In the RHS, reference waves generated by the feeds are injected into the waveguide directly. The waveguide then guides the reference wave to propagate on it.

\item \emph{Metamaterial Radiation Element}: The metamaterial radiation element is made of artificial composite material with supernormal electromagnetic properties or structures, whose electromagnetic response can be intelligently controlled by the magnetic or electric bias field. Each element is excited by the reference wave such that the radiation characteristic of the reference wave is dictated by the electromagnetic response of each radiation element and the leaky complex propagation constant.

\end{itemize}


\subsection{Holographic Principle}\label{IIB}
For the RHS, holographic beamforming is realized by constructing the \emph{holographic pattern}, which records the interference between the reference wave and the object wave according to the holographic interference principle. Utilizing the holographic pattern, the RHS can effectively control the radiation amplitude of the reference wave to obtain the desired radiation directions.

Below we further elaborate on the holographic principle. As shown in Fig. \ref{fig1}, the considered RHS has $K$ feeds and $M\times N$ radiation elements. At the $(m,n)$-th radiation element, the phase of the object wave $\Psi_{obj}$ propagating in the direction $(\theta_0, \varphi_0)$ is determined by the product of the propagation vector in free space $\mathbf{k}_f$ and the position of the $(m,n)$-th radiation element $\mathbf{r}_{m,n}$. The phase of the reference wave $\Psi_{ref}$ generated by feed $k$ is determined by the product of the propagation vector of the reference wave $\mathbf{k}_s$ and the distance vector from feed $k$ to the $(m,n)$-th radiation element $\mathbf{r}_{m,n}^k$.

According to the holographic interference principle, the interference $\Psi_{intf}$ between the reference wave and the object wave is  $\Psi_{obj}\Psi_{ref}^{\ast}$. When such an interferogram is excited by the reference wave, the generated beam satisfies
$\Psi_{intf}\Psi_{ref}\propto \Psi_{obj}|\Psi_{ref}|^2$,
such that the wave propagating in the direction $(\theta_0, \varphi_0)$ is generated.

To embody the whole surface interferogram (i.e., $\Psi_{intf}=\Psi_{obj}\Psi_{ref}^{\ast}$), the RHS adopts an amplitude-controlled method to construct the holographic pattern by controlling the radiation amplitude of the reference wave at each radiation element instead of phase-controlled beamforming. Specifically, each radiation element is tuned electrically to resonate at a frequency and radiate the reference wave. The radiation elements whose radiated waves are in phase at the desired directional beam  are tuned to radiate strongly, while the radiation elements that are out of phase are detuned so as not to radiate\cite{MSN-2016}.

Note that the real part of the interference (i.e., $\text{Re}[\Psi_{intf}]$) is the cosine value of the phase difference between the object wave and the reference wave. The value of $\text{Re}[\Psi_{intf}]$ decreases as the phase difference becomes larger, which exactly meets the amplitude control requirements, such that $\text{Re}[\Psi_{intf}]$ can represent the radiation amplitude of each radiation element. To avoid negative values, $\text{Re}[\Psi_{intf}]$ is normalized to $[0, 1]$. The radiation amplitude of each radiation element to generate the wave propagating in the direction $(\theta_0, \varphi_0)$ can then be parameterized mathematically by
\begin{equation}\label{M}
\mathrm{M}(\mathbf{r}_{m,n}^k,\theta_0, \varphi_0)=\frac{\text{Re}[\Psi_{intf}(\mathbf{r}_{m,n}^k, \theta_0, \varphi_0)]+1}{2}.
\end{equation}
This provides the basic principles for holographic beamforming based on amplitude controlling, which is different from the traditional phase-controlled beamforming.

\section{Fabrication Methodologies and Full-wave analysis of Reconfigurable Holographic Surfaces}

\subsection{Fabrication Methodologies}\label{IIIA}
The key issue in the fabrication of the RHS is to design metamaterial radiation elements with controllable radiation amplitude. There are three typical fabrication methodologies as introduced below.

\begin{figure}[t]
\centering
\includegraphics[width=3.5in]{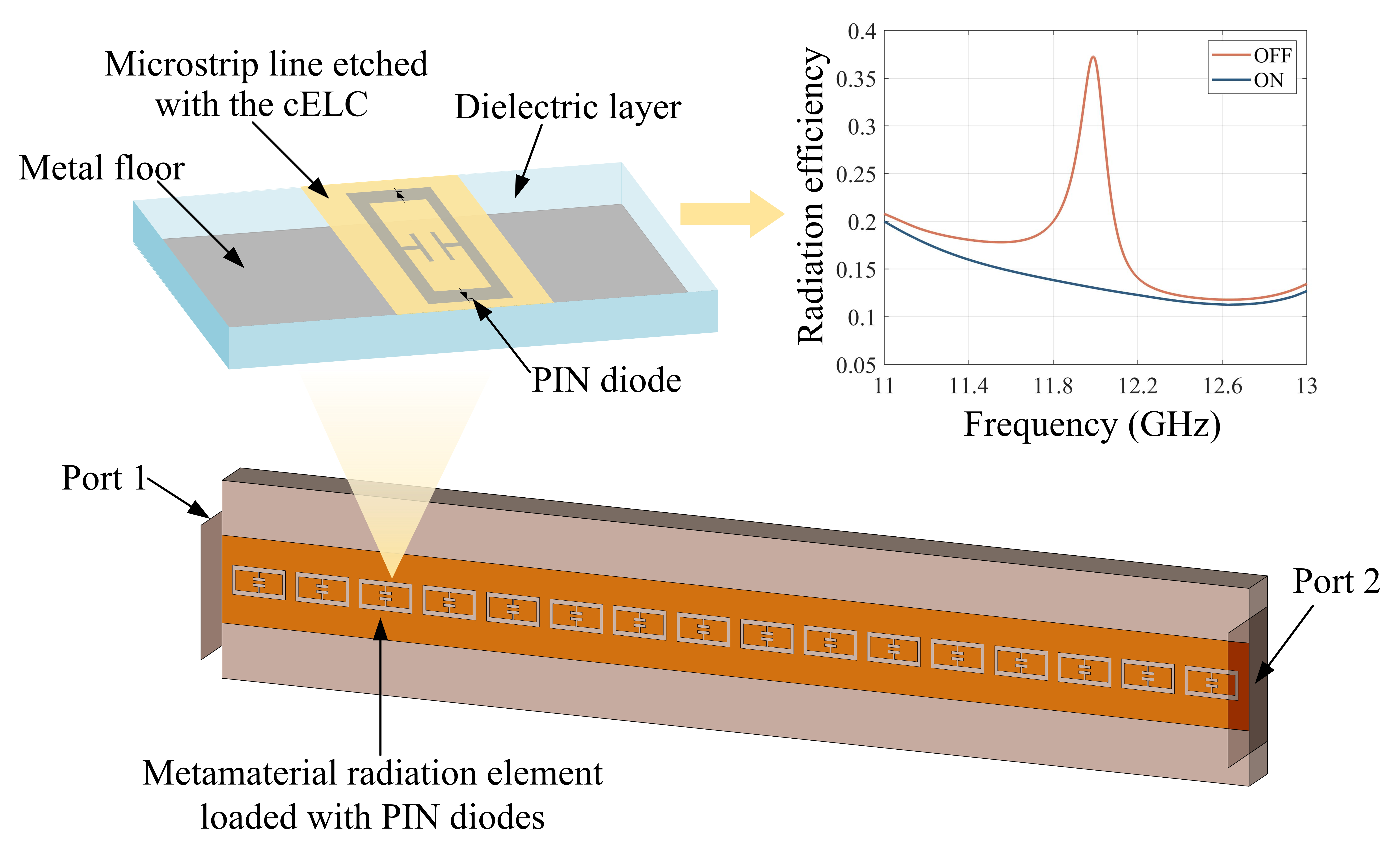}
\caption{Illustration of the metamaterial radiation element and the RHS simulated in the CST software.}
\label{figa}
\end{figure}

\subsubsection{PIN Diodes}
To design the tunable metamaterial radiation element with PIN diodes, a complementary electric-LC (cELC) resonator connected with PIN diodes\footnote{There are also other types of tunable metamaterial radiation elements such as slots and complementary
meander lines \cite{DO-2016}.} has been proposed in\cite{DO-2016} as shown in the upper half of Fig. \ref{figa}. The resonant frequency and radiation efficiency of the element can be modified by adjusting the cELC's geometric properties.

The mutual inductance of the cELC can be changed by applying biased voltage to the PIN diode to control its state (ON/OFF). Specifically, the radiation amplitude of each element can be calculated by (\ref{M}) for a given beam direction. If the amplitude value of an element is larger than a predefined threshold value, the PIN diode is in the OFF state and the element will radiate energy into the free space. Otherwise, the PIN diode is in the ON state and energy is barely radiated into the free space.

\subsubsection{Varactor Diodes \& Liquid Crystals}
To cover the shortage of PIN diodes which can only achieve discrete amplitude control, cELC resonators utilizing varactor diodes or liquid crystals to replace PIN diodes have been designed \cite{T-2016}. Since the capacitance of varactor diodes and liquid crystals can be changed by different biased voltage, the mutual inductance of the cELC and the radiation amplitude of the element can be changed continuously. The unique advantage of liquid crystals is that they show linear properties and work well in high-frequency bands, overcoming the nonlinearities of PIN diodes and varactor diodes caused by the increasing parasitic resistance in high-frequency bands\cite{CM-2010}.


\vspace{-0.2cm}
\subsection{Full-Wave Anlysis}
According to the basics of the RHS and its methodologies, we utilize the CST software to conduct a full-wave analysis of a one-dimensional RHS\footnote{The one-dimensional RHS can be extended to a two-dimensional RHS, where the radiation elements are embedded in a two-dimensional waveguide. For simplicity, we adopt a one-dimensional RHS for full-wave analysis to briefly illustrate the RHS's capability of accurate beam steering based on holographic beamforming.}. As shown in the lower half of Fig. \ref{figa}, the RHS consists of 16 cELC-based metamaterial radiation elements. The RHS is fed from port 1 and connected with a matching load at port 2. The reference wave generated by the feed in port 1 continuously radiates energy from the radiation element in the process of propagating from port 1 to port 2 along the microstrip line. The residual energy of the reference wave is absorbed by port 2.

Fig. 3 shows the radiation efficiency of each radiation element (i.e., the percentage of the energy radiated out of each radiation element to the total energy fed to it) in the ON and OFF states of the PIN diode. It can be seen that at the element's resonant frequency 12 GHz, the radiation efficiency of the element is about 37\% when the PIN diode is in the OFF state, indicating that the element will radiate much energy of the reference wave into free space. In contrast, when the PIN diode is in the ON state, the radiation efficiency is about 13\% such that little energy will be radiated\footnote{For an RHS with 16 radiation elements, considering the number of elements with PIN diodes in ON/OFF state for different object beams, the radiation efficiency of the element with PIN diodes in the OFF and ON state is required to be 30\%-40\% and lower than 15\%, respectively\cite{T-2016}. Under this requirement, when the input power is 0.5W, most of the incoming energy can be radiated into the free space.}.

\begin{figure}[t]
\centering
\includegraphics[width=2.8in]{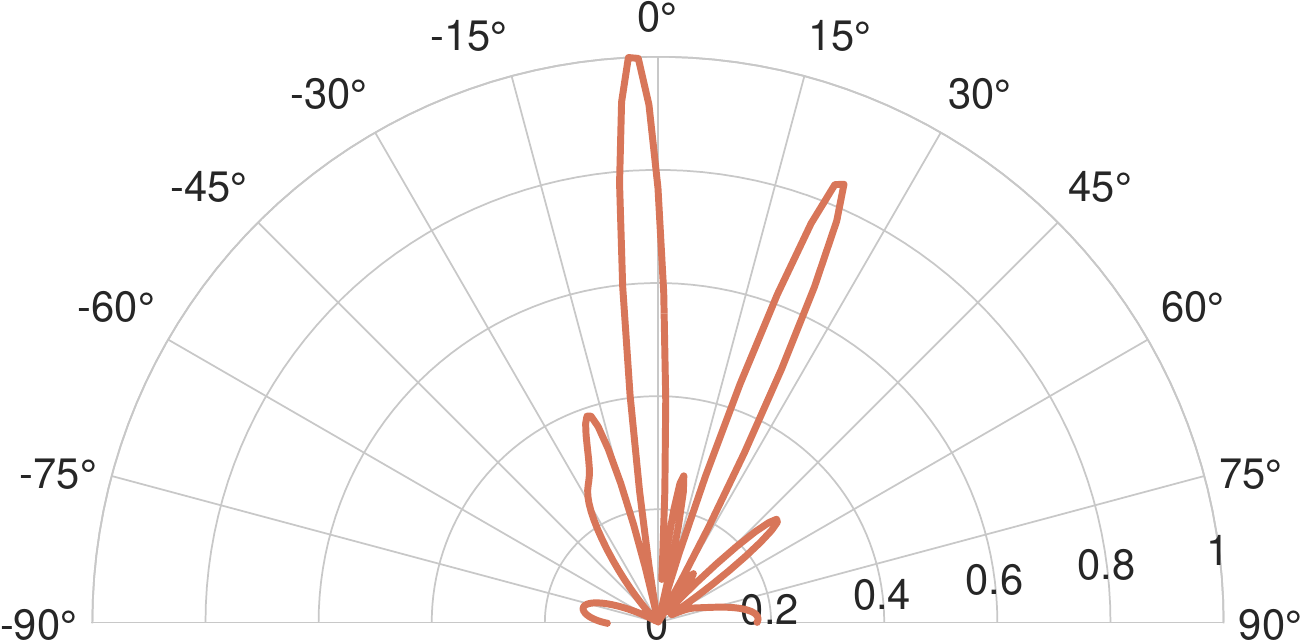}
\caption{Radiation pattern of the one-dimensional RHS.}
\label{single}
\end{figure}

Fig. \ref{single} shows the radiation pattern of this one-dimensional RHS when the desired beam directions are $-3^{\circ}$ and $23^{\circ}$, where the state of the PIN diodes in each radiation element is determined by a weighted summation of (1) corresponding to each object beam. Fig. 4 demonstrates that the ON and OFF states of the PIN diodes lead to different radiation amplitude at each element, thereby changing the beam direction.

\begin{figure*}[t]
\centering
\includegraphics[width=6.5in]{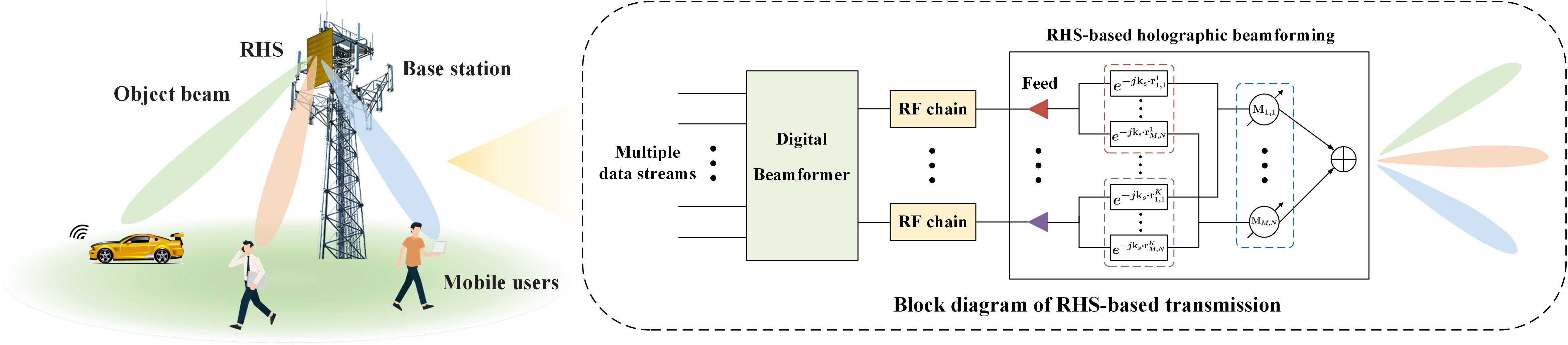}
\caption{RHS-aided multi-user communication system.}
\label{system_model1}
\end{figure*}

\section{RHS for Wireless communications}
In this section, we consider RHS-aided wireless communications, where a hybrid beamforming scheme is developed for sum rate maximization.

\subsection{RHS-Aided Wireless Communications}


Consider a downlink multi-user communication system where a BS equipped with an RHS transmits data streams to $L$  users and each user requires a single data stream from the BS. To maximize the sum rate, we propose a hybrid beamforming scheme for RHS-aided multi-user communication systems. Specifically, the BS performs the signal processing at the baseband since the RHS does not have any digital processing capability. The RHS then performs holographic beamforming to transmit the signal to each user.

As shown in Fig. \ref{system_model1}, the BS first encodes $L$ different data streams via a digital beamformer $\mathbf{V}\in \mathds{C}^{K\times L}$ at baseband and then up-converts the processed signals to the carrier frequency by passing through $K$ RF chains. Specifically, each RF chain is connected with a feed of the RHS, and each RF chain first up-converts transmitted signals to the carrier frequency and then sends the up-converted signals to its connected feed. The feed then transforms the high-frequency current into an electromagnetic wave, which is also called reference wave, propagating on the RHS. The reference wave will be transformed into a leaky wave through radiation elements on the RHS and leaks out its energy into free space for radiation, where the radiation amplitude of the reference wave at each radiation element is controlled via a holographic beamformer $\mathbf{M}\in \mathds{C}^{MN\times K}$ to generate desired directional beams.


\vspace{-0.2cm}
\subsection{Sum Rate Optimization}
We aim to maximize the sum rate by jointly optimizing the digital beamformer $\mathbf{V}$ and the holographic beamformer $\mathbf{M}$ subject to the transmit power constraint and the normalized radiation amplitude constraint. This is a non-trivial task since the traditional phase-controlled analog beamforming scheme is not applicable to amplitude-controlled holographic beamforming. A new beamforming scheme needs to be developed to handle the complex-domain optimization problem subject to the unconventional real-domain amplitude constraints, coupled with the superposition of the radiation waves from different radiation elements.

To tackle the above challenges, we first decompose the sum rate maximization problem into two subproblems, i.e., the digital beamforming subproblem subject to the transmit power constraint and the holographic beamforming subproblem subject to the normalized radiation amplitude constraint. We then develop a joint sum rate optimization algorithm to solve the digital beamforming subproblem and the holographic beamforming subproblem in an iterative manner. Two algorithms to solve these two subproblems are elaborated below.
\subsubsection{Digital Beamforming}
The digital beamforming subproblem can be solved by ZF beamforming together with power allocation for interference alleviation among users \cite{BHL-2020}.
\subsubsection{Holographic Beamforming}
The key idea to solve the holographic beamforming problem is to recast the sum-rate expression into the maxima of a concave function via
the fractional programming technique \cite{RBS-2020}. Specifically, we utilize the affine characteristic of linear functions to eliminate the non-convex terms in the expression of the signal to interference plus noise ratio (SINR) of each user. The modulus of the inner product of a real-valued matrix with other matrices in the expression of the SINR can be well approximated by a linear function of its elements. The holographic beamforming optimization problem is then reformulated as a concave optimization problem via the fractional programming technique. Finally, by adopting the Lagrangian dual to relax the normalized radiation amplitude constraint, the closed-form optimal solution of the holographic beamformer can be derived by taking the derivatives of the Lagrangian.

In summary, in each iteration of the sum rate maximization algorithm, the two subproblems are solved alternatively. The iterations are completed until the value difference of the sum rate between two adjacent iterations is less than a predefined threshold.


\subsection{Performance Evaluation}
\begin{figure}[t]
\centering
\includegraphics[width=3in]{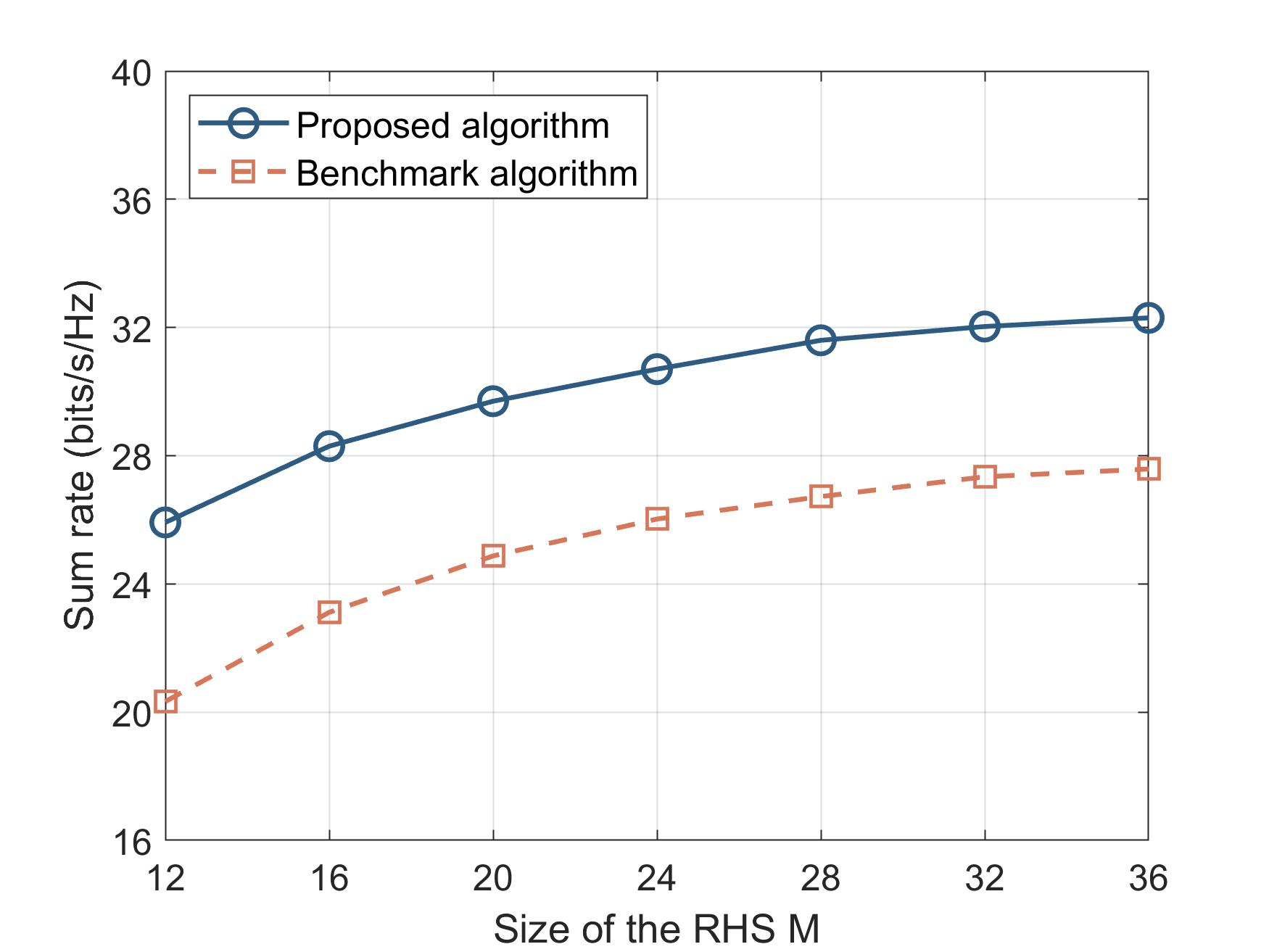}
\caption{Sum rate versus the size of the RHS.}
\vspace{-0.2cm}
\label{size}
\end{figure}

Simulated performance results are obtained using MATLAB software. Simulation parameters are set based on the 3GPP specifications \cite{3GPP}.
Fig. \ref{size} illustrates the sum rate versus the size of the RHS\footnote{For convenience, we set $M=N$ and adopt $M$ to represent the size of the RHS.}, i.e., $M$. It can be seen that the sum rate increases rapidly with a small value of $M$ and gradually flattens as $M$ continues to grow. The main reason is that the increment of $M$ first contributes to space multiplexing gain, and then only contributes to power gain due to the high correlation between different channel links. Moreover, it can be seen that the proposed joint optimization algorithm outperforms the benchmark algorithm, where the ZF digital beamforming is first performed, and the holographic beamformer is a direct superposition of the radiation amplitude distribution corresponding to each object beam according to (\ref{M}). This indicates the effectiveness of our proposed algorithm for sum rate maximization.

\section{Key Challenges in RHS-aided wireless communications}
Although RHSs have unique advantages over conventional antenna technologies, several key challenges remain to be solved in RHS-aided wireless communications.


\vspace{-0.2cm}
\subsection{Fundamental Design}

\subsubsection{Size Design}
In general, a larger-sized RHS can provide a higher antenna gain. However, the energy of the reference wave is steadily reduced with increasing propagation distance. If an RHS is oversized, the remaining energy is very weak when the wave propagates to the edge of the RHS such that the radiation elements at the edge of the RHS are redundant. Therefore, a moderate size design for an RHS is important.

\subsubsection{Element Spacing}
In theory, an RHS with smaller element spacing can generate a narrower and more accurate directional beam. However, in practical engineering, the mutual coupling effect between elements becomes stronger as the element spacing decreases, such that the radiation characteristics of the RHS will worsen. Therefore, the element spacing of the RHS should be carefully designed.



\subsection{Transmission Scheme Design}

\subsubsection{Joint Transceiver Design}
Since the RHS supports two-way communications, the joint transceiver design needs to be considered for capacity enhancement. Different from the conventional RF combiner relying on phase shifters at the receiver, the RHS modulates the incident plane wave from free space based on the holographic pattern, such that a novel receive beamforming scheme needs to be developed for receiving efficiency improvement.

\subsubsection{Channel Estimation}
Since the RHS is composed of numerous closely spaced radiation elements, the overhead required for channel estimation will be overwhelming due to pilot training and channel state information feedback. To reduce the pilot training overhead and enable a fast and accurate channel estimation, the pilot beam pattern, which is coupled with channel characteristics of the RHS-aided transmission needs to be carefully designed.

\section{Conclusions}
In this article, we have considered the RHS for future 6G communication networks where the RHS can achieve accurate beam steering with low power consumption and hardware cost by leveraging the holographic technique. The basics of the RHS including its hardware structure and holographic principle have been introduced, based on which its fabrication methodologies and full-wave analysis have been presented. Notably, we have proposed a hybrid beamforming scheme for RHS-aided wireless communications, where the digital beamforming performed at the BS and the holographic beamforming performed at the RHS are jointly optimized. Simulation results have indicated the effectiveness of the proposed hybrid beamforming scheme for sum rate maximization in multi-user communication systems. The key challenges in RHS-aided wireless communications have also been elaborated.

%

\end{document}